\documentclass[twocolumn,prb,showpacs,superscriptaddress,amsmath,amssymb]{revtex4}

\usepackage{graphicx}
\usepackage{amsmath}
\usepackage{color}

\def\AA{${\buildrel _{\circ} \over {\mathrm{A}}} $}

\begin{document}

\title{First-principles investigation of hyperfine interactions \\
for nuclear spin entanglement in photo-excited fullerenes}

\author{Vasileia Filidou}
\affiliation{Department of Materials, University of Oxford, Parks Road, 
Oxford OX1 3PH, United Kingdom}
\affiliation {CAESR, Clarendon Laboratory, Oxford University, Oxford,
OX1 3PU, United Kingdom}

\author{Davide Ceresoli}
\affiliation{Department of Materials, University of Oxford, Parks Road, 
Oxford OX1 3PH, United Kingdom}
\affiliation{CNR-ISTM, via Golgi 19, 20133 Milan, Italy}

\author{John J. L. Morton}
\affiliation{Department of Materials, University of Oxford, Parks Road, 
Oxford OX1 3PH, United Kingdom}
\affiliation {CAESR, Clarendon Laboratory, Oxford University, Oxford,
OX1 3PU, United Kingdom}

\author{Feliciano Giustino}
\affiliation{Department of Materials, University of Oxford, Parks Road, 
Oxford OX1 3PH, United Kingdom}
\email{vasileia.filidou@materials.ox.ac.uk}

\date{\today}

\begin{abstract}
The study of hyperfine interactions in optically excited fullerenes has recently 
acquired importance within the context of nuclear spin entanglement for quantum 
information technology. We here report a first-principles pseudopotential study 
of the hyperfine coupling parameters of optically excited fullerene derivatives 
as well as small organic radicals. The calculations are performed within the 
gauge-invariant projector-augmented wave method [C. Pickard and F. Mauri, 
Phys.\ Rev.\ B.\ {\bf 63}, 245101 (2001)]. In order to establish the accuracy 
of this methodology we compare our results with all-electron calculations 
and with experiment. In the case of fullerene derivatives we study the hyperfine 
coupling in the spin-triplet exciton state and compare our calculations with 
recent electron paramagnetic resonance measurements [M. Schaffry {\it et al.}, 
Phys.\ Rev.\ Lett.\ {\bf104}, 200501 (2010)]. We discuss our results in light 
of a recent proposal for entangling remote nuclear spins in photo-excited chromophores.
\end{abstract}

\pacs{33.15.Pw, 
      31.30.Gs, 
      71.20.Tx  
}

\maketitle

\section{Introduction}

In the area of quantum information processing substantial efforts are being devoted 
to exploit the magnetic properties of molecular systems in order to realize quantum 
logic operations.\cite{morton} In a recent proposal for entangling two nuclear spins 
using a photo-excited transient electron spin,\cite{marcus} the strength of the 
hyperfine interaction (HFI) has been identified as a key parameter determining 
the entanglement efficiency. These investigations raise the question on how to optimize 
such parameters, and how to design new materials with tailored magnetic 
properties for specific quantum logic operations.

First-principles calculations based on density-functional theory (DFT) have recently 
met with a number of successes in the calculation of the magnetic properties 
of materials, in particular when it comes to the building blocks of the spin 
Hamiltonian such as chemical shifts,\cite{shift,shift2,shift3,shift4,shield,shield2} 
J-couplings,\cite{tens,tens2} and g-tensors.\cite{gten,gten2} As magnetic spin 
resonance experimental techniques become more accurate, and novel applications 
to physics, biology, and materials science are being explored,\cite{bio,mat} 
DFT-based modelling of charge/spin interactions hold the promise of becoming 
a key player in the interpretation of measured magnetic responses and in the design 
of materials with tailored functionality.

In this work we aim to establish the validity and accuracy of DFT-based 
modelling in the calculation of HFI parameters in a class of materials of interest 
for quantum information technology. We here focus on the HFI interactions 
in the fullerene derivatives synthesized in Ref.~\onlinecite{marcus}, 
since electron spin resonance measurements on these materials have already
been performed.

The manuscript is organized as follows. In Sec.~\ref{sec.qip} we briefly review 
the connection between the HFI parameters and the efficiency of the quantum 
entanglement operations introduced in Ref.~\onlinecite{marcus}. 
In Sec.~\ref{sec.computation} we describe the computational methodology that 
we employ for calculating hyperfine tensors. In Sec.~\ref{sec.results.a}
we discuss our HFI calculations for a series of small radicals. In this section
we establish the accuracy and reliability of pseudopotential (PP) calculations
of the HFI by comparing with all-electron (AE) calculations and with experiment.
In Sec.~\ref{sec.results.b} we present our calculations of the HFI on photo-excited
fullerene derivatives, and we discuss our results in light of the 
entanglement protocol proposed in Ref.~\onlinecite{marcus}. In Sec.~\ref{sec.conc}
we present our conclusions.

\section{Entangling nuclear spins via the hyperfine interaction}\label{sec.qip}

In Ref.~\onlinecite{marcus} a quantum information processing protocol involving
nuclear spins has been introduced. This protocol involves the interaction 
of two nuclear spins attached to a fullerene molecule through functional additions
carrying isotopically labeled nuclei. The ground state of this fullerene
derivative is spin-unpolarized, and hence the HFI of the nuclear spins is
vanishing. Upon optical excitation a spin-triplet excitonic state is generated
in the molecule, and the nuclear spins interact with the spin-polarized 
electrons leading to a non-vanishing hyperfine coupling. As a result the
two nuclear spins experience an interaction which is mediated by the spin
of the exciton. Such interaction can be used to perform an entanglement
operation between the two nuclear spins. When the exciton decays and the
fullerene reverts to its spin-unpolarized ground state, the nuclear spins
may remain entangled.
The efficiency of this protocol results from the competition between three timescales: 
(i) the time required to perform the entanglement operation, (ii) the exciton lifetime, 
and (iii) the decoherence time of the entangled state.\cite{marcus} The competition 
between these timescales defines an optimum range of HFI strength for achieving maximum 
entanglement.

In order to establish the feasibility of the protocol proposed in Ref.~\onlinecite{marcus},
pulsed electron paramagnetic resonance experiments were performed on a prototype system, 
the fullerene diethyl malonate molecule [C$_{60}$-C$_7$H$_{12}$O$_4$, system M1 
in Fig.~\ref{fig1}(a)]. Although this system carries only one isotopically labeled
nuclear spin associated with the $^{13}$C atom of the functional group 
directly bonded to the cage, it was possible to extract relevant 
information such as the lifetimes of the exciton states and the HFI parameters.
For the system considered the measured relaxation rates of 90-100 $\mu$s
at 20-50 K correspond to maximum entangling power when the HFI parameters 
are in the range 2 MHz to 12 MHz, with an optimal value of 6 MHz.\cite{marcus}
The measured HFI parameter of 3~MHz in this prototype system with a single nuclear 
spin falls within the optimum theoretical range and motivates further work in this area.

The key open question which remains is how to identify systems involving multiple 
nuclear spins and exhibiting optimal HFI strengths. One possible route for answering
such question is to employ first-principles materials modelling based on 
density-functional theory in order to screen candidate systems prior to their 
experimental synthesis. Accordingly, in the present work we aim at establishing 
the accuracy of such first-principles calculations, in view of optimizing and 
designing functional materials for spin-based quantum information processing.

\section{Computational methodology}\label{sec.computation}

\subsection{Hyperfine tensor from first principles}

The coupling of the electron spin $\vec S$ to a set of $N$ nuclear spins 
$\vec I_l$ ($l=1,\dots N$) can be described using the following hyperfine 
Hamiltonian:\cite{blochl}
  \begin{equation}
  \hat{H} = \sum_{l=1,N} {\vec S}^{\,T}{\bf A}^{(l)}{\vec I}_l,
  \end{equation}
${\bf A}^{(l)}$ being the hyperfine coupling tensor associated with the $l$-th
nucleus at the site ${\bf R}_l$.
In atomic units the hyperfine tensor can be written as:\cite{blochl}
$A^{(l)}_{ij} = a^{(l)}\delta_{ij} + b^{(l)}_{ij}$,
with 
  \begin{equation}
  \label{eq:1}
  a^{(l)}  = \frac{{8\pi }} {3} g_e\mu_e g_l\mu_N \rho_{\rm s}({\bf R}_l),
  \end{equation}
and
  \begin{equation}
  \label{eq:2}
  b^{(l)}_{ij} = g_e\mu_e g_l\mu_N
  \int d{\bf r} \frac{{3 r_i r_j  - r^2\delta_{ij}}} {{r^5}}
  \rho_{\rm s}({\bf r}).
  \end{equation}
In these equations $\rho_{\rm s}({\bf r}) = 
\rho_\uparrow({\bf r})-\rho_\downarrow({\bf r})$ is the electron spin density,
$g_e$ the free-electron g-factor, $\mu_e$ the Bohr magneton, $g_l$  
the gyromagnetic ratio of the $l$-th nucleus and  $\mu_N$ the nuclear magneton. 
In Eq.~(\ref{eq:2}) ${\bf r}$ is relative to the nuclear site ${\bf R}_l$ 
and $r=|{\bf r}|$.
The component $a^{(l)}$ of the tensor ${\bf A}^{(l)}$ provides the isotropic
hyperfine interaction and is referred to as the Fermi contact term.
The traceless tensor $b^{(l)}_{ij}$ gives the anisotropic or dipolar
hyperfine interaction.
The anisotropic tensor $b^{(l)}_{ij}$ is conventionally given in terms 
of its eigenvaules $b^{(l)}_{xx}$, $b^{(l)}_{yy}$, and $b^{(l)}_{zz}$. We note
that the principal axis of the hyperfine tensor ${\bf A}^{(l)}$ and of its
anisotropic part $b^{(l)}_{ij}$ do coincide by construction.

The evaluation of the HFI parameters using Eqs.~(\ref{eq:1}) and (\ref{eq:2})
requires the calculation of the electron spin density. In the
dipole-dipole term $b^{(l)}$ the electron spin-density is weighted by the
factor $r^{-3}$ and the contributions outside of the core region are dominant.
In the Fermi contact term the electron spin density must be evaluated
at the position of the nucleus, therefore an accurate description 
of the electron density within the core region is essential.
While standard pseudopotential methods can be used for calculating 
the dipole-dipole term $b^{(l)}$, the calculation of the Fermi contact
contribution requires the use of all-electron methods or the projected 
augmented wave (PAW) method.\cite{blochl} 
The PAW method enables the reconstruction of the all-electron wavefunction
inside the core region, starting from a pseudopotential calculation of the
valence electronic structure, and has successfully been applied to the
calculation of hyperfine parameters.\cite{vdw}

\subsection{Systems considered}\label{sec.systems}

\begin{figure}[t!]
  \begin{center}
  \includegraphics[width=0.9\columnwidth]{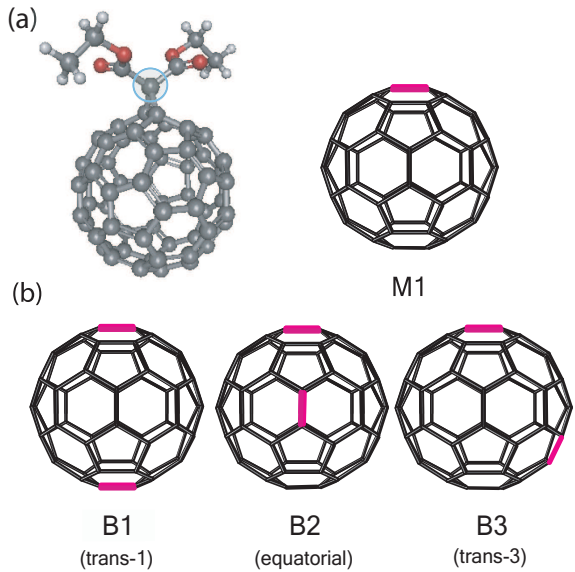}
  \caption{\label{fig1}
  (a) Ball-and-stick (left) and schematic (right) representations 
  of the diethyl malonate (C$_7$H$_{12}$O$_4$) C$_{60}$ mono-adduct (M1). 
  The bridging $^{13}$C atom carrying a nuclear spin is indicated.
  In the schematic representation the red bond indicates the anchoring point 
  of the adduct on the fullerene cage. Atomic color code: C (grey), O (red),
  H (white).
  (b) Schematic representations of three regioisomeric diethyl malonate 
  fullerene bis-adducts (B1-B3).
  }
  \end{center}
  \end{figure}

In Ref.~\onlinecite{marcus} we reported a preliminary calculation
of the isotropic component of the HFI tensor of the M1 system [Fig.~\ref{fig1}(a)],
and of a related fullerene derivative carrying two $^{13}$C nuclear spins
[system B1 in Fig.~\ref{fig1}(b)].
Here we complement the work of Ref.~\onlinecite{marcus} by calculating the
entire HFI tensor of these systems, and we further extend it to fullerene
derivatives of various conformations. Furthermore, we perform
calculations on a number of small open-shell molecules, both within PP and AE methods,
in order to establish the accuracy of the computational methodology.

We consider following small molecules containing hydrogen and oxygen: 
CH$_2$CH$_3$, CH$_2$CH, CO$_3$, H$_3$CO, HCO, H$_2$CO$^+$, and CH$_3$. 
In the case of diethyl malonate fullerene derivatives we consider 
the mono-adduct M1 [Fig.~\ref{fig1}(a)], as well as 
the three regioisomeric bis-adducts B1-B3 shown in Fig.~\ref{fig1}(b).
During the synthesis of stable regioisomeric bis-adducts 
from diethyl malonate and fullerene molecules, 
the trans-1 isomer [model B1 in Fig.~\ref{fig1}(b)] is produced with
the lowest yield (0.8\%), while the equatorial and the trans-3
isomers [models B2 and B3 in Fig.~\ref{fig1}(b), respectively]
are produced with the highest yields (16\% and 12\%, respectively).\cite{hirsh}

\subsection{Calculations details}\label{sec.calc-details}

We perform pseudopotential (PP) density-functional theory calculations within the 
generalized gradient approximation of Perdew, Burke, and Ernzerhof (PBE).\cite{gga} 
We describe the valence electronic wavefunctions using a planewaves basis set
with a kinetic energy cutoff of 70 Ry. We take into account the core-valence 
interaction using Troulier-Martins normconserving pseudopotentials.\cite{tm,fuchs}
For the calculation of the HFI tensor we use the gauge-including projector-augmented
wave method (GIPAW) of Ref.~\onlinecite{gipaw}.  We use periodic simulation cells,
and in order to minimize interactions between periodic replicas of the molecules, 
we use cells of size 12 \AA\ for the small molecules, and cells of size up to 
24 \AA\ for the fullerene derivatives. All PP calculations are performed 
using the {\tt quantum-ESPRESSO} software package.\cite{qe} 

For the calculation of the Fermi contact term, the accurate description of the 
electronic wavefunctions in the core region is required. In pseudopotential methods
only valence electrons are described, and in some cases this might lead to
significant errors due to the neglect of core polarization effects. 
Core polarization refers to the spin polarization of core electrons induced by
the spin of valence electrons, and leads to an additional contribution to
the Fermi contact term.\cite{FW,WF} Within PP calculations we introduce core
polarization corrections using the perturbative approach of Ref.~\onlinecite{corerelax}
as follows. For each atom with core electrons (C and O) we calculate the spherical 
average of the spin density around the nuclues up to a radius of 5 a.u. using
a logarithmic radial grid. This average is used to evaluate the perturbation $\Delta V$
due to valence polarization following Eq.~(20) of Ref.~\onlinecite{corerelax},
and subsequently the contribution to the Fermi contact term arising from
the core polarization using Eq.~(17) of Ref.~\onlinecite{corerelax}.
The atomic all-electron wavefunctions are stored in the pseudopotentials files
and the integrations are carried out on the radial grid.
In the following, pseudopotential calculations including the core polarization 
correction will be referred as ``PP-CP'', while those without core polarization 
will be denoted simply by ``PP''.
In the case of light atoms such as those considered in this work relativistic effects
are expected to be negligible, therefore we generated the pseudopotentials 
by solving the non-relativistic Schr\"odinger equation (it is custom practice 
to solve the scalar-relativistic equation only for atoms starting from the
fourth row of the periodic table).
In evaluating the Fermi contact term, the first relativistic correction
can be taken into account through a spherical integration of the spin density
up to the Thomson radius~$r_T$, which provides an estimate
of the finite size of the nucleus. However, the Thomson radii of
H and C fall around the first point of our pseudopotential radial grid, 
therefore it seems more appropriate to evaluate the spin density 
at the nucleus (this is done by extrapolating the three innermost values on the grid).

In order to establish the accuracy of PP calculations of HFI parameters we also
perform AE calculations using the PBE exchange and correlation, as implemented 
in the {\tt Gaussian 09} program package.\cite{gau} For the AE calculation of the HFI 
tensor we adopt the EPR-II basis, which has been purposely designed 
to accurately describe the Fermi contact term.\cite{eprII}

In order to study the HFI in optically-excited fullerene derivatives we perform
calculations using the ground state of the $S_z=1$ electron spin configuration. 
In fact, while density-functional theory does not describe correctly excited states, 
the lowest spin-triplet state of C$_{60}$, with the HOMO and LUMO states singly 
occupied with parallel spins, is the ground state of the $S_z=1$ spin configuration, 
hence the use of DFT is legitimate.\cite{trip1,trip2}  

For each structure M1-B3 shown in Fig.~\ref{fig1} we optimize several initial 
geometries in both the singlet $S_z=0$ and triplet $S_z=1$ configurations
in order to identify the corresponding ground state. The 
energy difference between the triplet and the singlet states ranges from 1.29 to 1.33 eV
across all the models considered. This result is in agreement with the measured 
excitation energy of 1.30 eV of the non-functionalized fullerene C$_{60}$.\cite{ext1,ext2} 
This suggests that our description of the lowest triplet exciton state 
is adequate.

\section{Results and Discussion}\label{sec.results}

\subsection{HFI parameters of small molecules}\label{sec.results.a}

  \begin{table}
  \caption{\label{table1}
  Fermi contact hyperfine coupling parameters of small molecules containing 
  C, and H. We report calculations using the GIPAW method without accounting for 
  core-polarization effects (PP), calculations within GIPAW including core-polarization 
  corrections (PP-CP), all-electron calculations (AE), and measured Fermi contact 
  parameters. The atoms are listed according to corresponding chemical formula.}
  \vspace{0.2cm}
  \begin{center}
  \begin{tabular}{l l l r r r r r r r l }
  \hline\hline
  &Molecule & Atom &  & \multicolumn{7}{c}{Fermi contact term (MHz)}\\
  &        &       &  &PP &  &PP-CP & &AE& \multicolumn{2}{c}{Expt.} \\
  \hline
  &CH${_2}$CH  &C(1) &  &397&  &305\hspace{.4cm}  & &306 &\hspace{0.4cm}302&$\!\!^{\rm a}$\\
  & & H(1) & &125 &  && &124 & 96& \\
  & &H(2) & &191 &   && &190 & 192& \\
  & &C(2) & &-27 &   &-17\hspace{.4cm}  & &-18 & -24&\\
  & & H(3) & & 38 &  && &43 & 37&\\
  \hline
  & CH${_2}$CH${_3}$ &C& &205& &81\hspace{.4cm}   & &79& 83&$\!\!^{\rm b}$ \\
  & & H& &-63 & && &-61 &-69& \\
  & & C & &-40 & &-38\hspace{.4cm}  & &-34& -38&\\
  & & H & &51 & && &40& 75& \\
  \hline
  &CH${_3}$ &C& &214& &79\hspace{.4cm}  & &72& 80&$\!\!^{\rm g}$\\
  & & H& &-69& && &-66& -69&\\
  \hline
  &H${_3}$CO &H & & 94& &  & & 95& 122 \\
  &          &C & &-35& &-44\hspace{.4cm}  & &-37& -44&$\!\!^{\rm d}$\\
  \hline
  &HCO &H & &379& && &367& 356&\\
  & &C& &435 & &428\hspace{.4cm}  & &389& 367&$\!\!^{\rm e}$\\ 
  \hline
  &H${_2}$CO$^{+}$ &H & &279& && &372& 372&\\
  & &C & &-76& &-83\hspace{.4cm}  & & -87& -109&$\!\!^{\rm f}$\\
  \hline
  &CO${_3}$ &C& &-31 & &-31\hspace{.4cm}  & &-28& -32&$\!\!^{\rm c}$\\
  \hline
  \hline
  \end{tabular}
  \end{center}
  \hspace{-4.8cm}
  $^{\rm a}$Refs.~\onlinecite{ch2ch13iso-a,ch2ch13iso-b},\\
  \hspace{-5.4cm}
  $^{\rm b}$Ref.~\onlinecite{co2iso},\\
  \hspace{-5.4cm}
  $^{\rm c}$Ref.~\onlinecite{co3iso},\\
  \hspace{-5.0cm}
  $^{\rm d}$Ref.~\onlinecite{h3coiso,h3co},\\
  \hspace{-5.4cm}
  $^{\rm e}$Ref.~\onlinecite{hcoiso-aniso},\\
  \hspace{-5.4cm}
  $^{\rm f}$Ref.~\onlinecite{h2coiso},\\
  \hspace{-5.4cm}
  $^{\rm g}$Ref.~\onlinecite{ch3iso}.
  \end{table}

  \begin{table*}
  \caption{\label{table2}
  Dipolar hyperfine coupling parameters of small molecules containing
  C and H. We report PP calculations, AE calculations, and measured parameters.
  The atoms are listed according to corresponding chemical formula.}
  \vspace{0.2cm}
  \begin{minipage}[l]{0.49\textwidth}
  \begin{flushright}
  \begin{tabular}{l c c r r r l }
  \hline
  \hline
   Molecule & Atom & \multicolumn{4}{c}{Dipolar HFI (MHz)} \\
  \hline
    & & Axis\hspace{0.2cm} & PP & AE & Exp.\ \\
  \hline
  CH${_2}$CH    &  &$b_{xx}$ & -86  & -72 &     &                \\
                &C(1) &$b_{yy}$ & -64  & -56 &     &                \\
                &  &$b_{zz}$ & 148  & 127 &     &                \\
                &  &$b_{xx}$ &  37  &  37 &  38 & $\!\!^{\rm h}$ \\
                &H(3) &$b_{yy}$ &  -6  &  -6 & -11 &                \\
                &  &$b_{zz}$ & -32  & -32 & -27 &                \\
                
  \hline                                                             
  CH$_2$CH$_3$  
                &  &$b_{xx}$ & -83  & -73 &     &                \\
                &C &$b_{yy}$ & -82  & -72 &     &                \\
                &  &$b_{zz}$ & 165  & 145 &     &                \\
                &  &$b_{xx}$ &  35  &  38 &  35 & $\!\!^{\rm i}$ \\
                &H &$b_{yy}$ &   2  &   0 &   1 &                \\
                &  &$b_{zz}$ & -38  & -38 & -36 &                \\
                &  &$b_{xx}$ &  -3  &  -3 &     &                \\
                &C &$b_{yy}$ &   1  &   1 &     &                \\
                &  &$b_{zz}$ &   3  &   2 &     &                \\
                &  &$b_{xx}$ &  -4  &  -4 &  -3 & $\!\!^{\rm i}$            \\
                &H &$b_{yy}$ &  -3  &  -3 &  -3 &                \\
                &  &$b_{zz}$ &   7  &   9 &   6 &                \\
  \hline 
  CH$_3$        &  &$b_{xx}$ & -70  & -74 & -63 & $\!\!^{\rm l}$ \\ 
                &C &$b_{yy}$ & -72  & -74 & -63 &                \\   
                &  &$b_{zz}$ & 150  & 148 & 127 &                \\   
                &  &$b_{xx}$ & -36  & -39 & -36 & $\!\!^{\rm m}$ \\  
                &H &$b_{yy}$ &  -1  &  -1 &   -1 &                \\ 
                &  &$b_{zz}$ &  35  &  39 &  35 &                \\ 
  \hline
  \hline
  \end{tabular}
  \end{flushright}
  \end{minipage}
  \begin{minipage}[r]{0.49\textwidth}
  \vspace{-1.4cm}
  \begin{flushleft}
  \hspace{0.2cm}
  \begin{tabular}{l c c r r r l }
  \hline
  \hline
  Molecule & Atom & \multicolumn{4}{c}{Dipolar HFI (MHz)} \\
  \hline
  & & Axis\hspace{0.2cm} & PW & AE & Exp.\ \\
  \hline
  H${_3}$CO     &  &$b_{xx}$ &   7  &   6 &     &                \\  
                &C &$b_{yy}$ &   2  &   2 &     &                \\  
                &  &$b_{zz}$ &  -9  &  -8 &     &                \\  
                &  &$b_{xx}$ &  -8  &  -7 &     &                \\         
                &H &$b_{yy}$ &  -4  &  -4 &     &                \\ 
                &  &$b_{zz}$ &  12  &  12 &     &                \\ 
  \hline                                                             
  HCO           &  &$b_{xx}$ & -43  & -42 & -48 & $\!\!^{\rm c}$ \\                        
                &C &$b_{yy}$ & -33  & -38 & -24 &                \\
                &  &$b_{zz}$ &  76  &  80 &  72 &                \\       
                &  &$b_{xx}$ & -14  & -14 & -14 &                \\
                &H &$b_{yy}$ &  -8  &  -7 &  -8 &                \\
                &  &$b_{zz}$ &  23  &  21 &  23 &                \\
  \hline                                                                                   
  H$_2$CO$^{+}$ &  &$b_{xx}$ & -25  & -21 & -15 & $\!\!^{\rm k}$ \\   
                &C &$b_{yy}$ &   7  &   6 &   4 &                \\
                &  &$b_{zz}$ &  18  &  16 &  10 &                \\ 
                &  &$b_{xx}$ & -10  & -10 &  -9 &                \\ 
                &H &$b_{yy}$ &  -8  &  -7 &   5 &                \\    
                &  &$b_{zz}$ &  18  &  18 &   4 &                \\   
  \hline                                                                                
  CO$_3$        &  &$b_{xx}$ &   5  &   4 &   4 & $\!\!^{\rm j}$ \\
                &C &$b_{yy}$ &   5  &   4 &   4 &                \\
                &  &$b_{zz}$ &  -9  &  -9 &  -8 &                \\
  \hline
  \hline
  \end{tabular}
  \end{flushleft}
  \end{minipage}
  \begin{center}
  \hspace{-1.5cm}
  $^{\rm h}$ Ref.~\onlinecite{ch2chaniso},
  $^{\rm i}$ Ref.~\onlinecite{ch2ch3aniso},
  $^{\rm j}$ Ref.~\onlinecite{co3iso},
  $^{\rm k}$ Ref.~\onlinecite{hcoiso-aniso},
  $^{\rm l}$ Ref.~\onlinecite{h2coaniso},
  $^{\rm m}$ Ref.~\onlinecite{ch3anisoC},
  $^{\rm n}$ Ref.~\onlinecite{ch3anisoH}.
  \end{center}
  \end{table*}

In Table \ref{table1} we present the results of our pseudopotential calculations 
and we compare with all-electron calculations and with experiment. 
For a quantitative analysis we present both pseudopotential results with (PP-CP) 
and without (PP) core polarization correction. 

In the cases of CH${_2}$CH, CH${_2}$CH$_3$, and CH${_3}$ Table \ref{table1} shows
that the neglect of core polarization for the C atoms leads to an underestimation 
of the Fermi contact term by as much as 92-135 MHz. When the core polarization 
correction is introduced, all the calculations tend to agree rather well with 
our AE results and with experiment. In the case of H atoms, where the core and
valence do coincide, our PP results are in good agreement with AE and experimental
data.

\begin{figure}
  \begin{center}
  \includegraphics[width=\columnwidth]{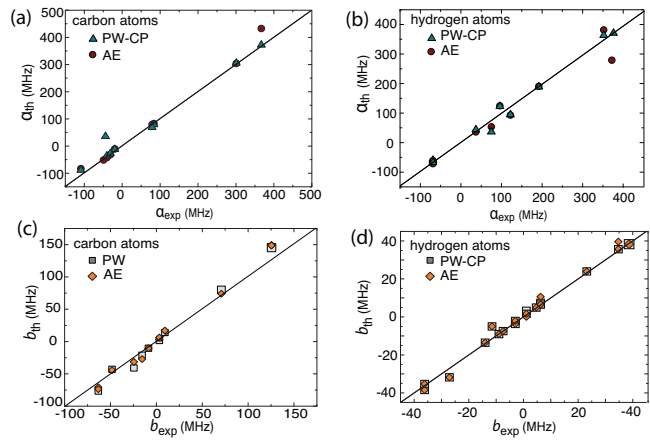}
  \caption{\label{fig2}
  Comparison between the HFI parameters calculated using PP-CP and AE methods
  and experimental data, from Tables \ref{table1} and \ref{table2}. 
  (a) Fermi contact term of C atoms. (b) Fermi contact 
  term of H atoms. (c) Dipolar coupling of C atoms. (d) Dipolar coupling of H atoms.
  If the calculations were able to predict 
  experimental data exactly, then all the data points would
  fall on the diagonal solid line.}
  \end {center}
  \end {figure}

Figures~\ref{fig2}(a),(b) show a comparison between the Fermi contact
term for C and H atoms calculated with the PP-CP method or the AE method,
and experiment. The relative accuracy of PP-CP calculations
and AE calculations in the case of C atoms appears similar for this 
set of test molecules. Indeed the largest deviations from experiment 
are of 17\% and 13\% for PP-CP and AE calculations, respectively.
In absolute terms the largest deviations from experiment are 61 MHz (22 MHz) for
C atoms using PP-CP (AE) calculations, and 93 MHz (28 MHz) for
H atoms using PP-CP (AE) calculations.

In Table \ref{table2} we present the results of calculated dipolar hyperfine couplings.
The same data set is reported in 
Figs.~\ref{fig2}(c),(d) for C and H atoms, respectively.
The largest absolute deviation between our PP calculations and experiment 
is 23 MHz, while the largest deviation for the case of AE calculations is 21 MHz. 
Again PP calculations and AE calculations appear to perform similarly,
and are both able to provide reasonably reliable data when compared to
experiment.

What we learn from these tests is that, on the one hand, 
in the case of the Fermi contact term, the inclusion of core polarization 
in PP calculations is critical 
to achieving agreement with AE calculations and with experiment.
On the other hand, PP calculations of the dipolar HFI interaction are
already in reasonable agreement with AE data and experiment.
In the case of HCO we observe that the deviation between the 
calculated isotropic HFI (either PP-CP or AE) and experiment 
can be rather large, hence care must be used in the interpretation
of the calculated HFI parameters. This discrepancy requires further investigation.

\subsection{HFI parameters of fullerene derivatives}\label{sec.results.b}

\begin{figure}[h!]
  \begin{center}
  \includegraphics[width=0.75\columnwidth]{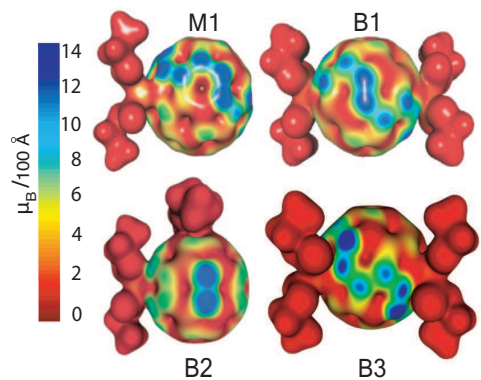}
  \caption{\label{fig3}
  Spin density distribution in the lowest triplet excited state of the four
  fullerene derivatives M1 and B1-3 considered in this work. The color code indicates
  the magnitude of the spin density (the density isosurface corresponds to
  a isovalue of 0.07 electron/\AA$^3$).
  }
  \end{center}
  \end{figure}

In this section we present the HFI parameters calculated for the fullerene derivatives 
of Fig.~\ref{fig1}.  Figure~\ref{fig3} shows the calculated spin density distribution 
for the structures M1 and B1-3 in the triplet state. The spin density is greatest 
(blue regions) on the fullerene cages, while it is rather small (red regions) 
on the adducts and the isotopically labeled C nuclei. As the strength of the HFI 
scales with the spin density [cf.\ Eqs.~(\ref{eq:1}),(\ref{eq:2})], Fig.~\ref{fig3} 
suggests that we should expect a rather small HFI strength for all the structures
considered.

This observation is consistent with the calculated HFI parameters reported in
Table \ref{table3}, which are of the order of only a few MHz. 
All the pseudopotential calculations in this table include 
the core polarization correction (PP-CP) discussed in Sec.~\ref{sec.calc-details}. 
For the case of the M1 structure both PP-CP and AE calculations are in
good agreement with the experimental data from Ref.~\onlinecite{marcus}.
In light of the error bars established in Sec.~\ref{sec.results.a} for the
test molecules, this level of agreement (around 1 MHz) may be somewhat fortuitous.
However it is reasonable to assume that, due to error cancellation, 
our calculations should provide a reliable description of the {\it relative} 
magnitude of the HFI parameters across various fullerene derivatives.
It is possible that accounting for thermal motion may lead to slightly 
different HFI parameters in our calculations, however we expect that such
effects will fall within our error bars established in Sec.~\ref{sec.results.a}.
Morover the effect of thermal motion should be similar across the
various derivatives, therefore the relative magnitude of the HFI parameters
is expected to be reliable.

  \begin{figure}[t!]
  \begin{center}
  \includegraphics[width=0.9\columnwidth]{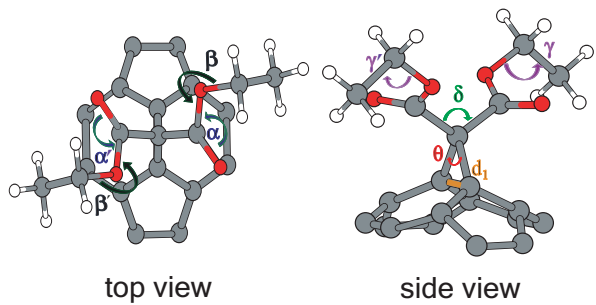}
  \caption{\label{fig4}
  Ball-and-stick representation of the diethyl malonate adduct on a portion
  of the C$_{60}$ fullerene (color code as in Fig.~\ref{fig1}).
  The angles in all the systems considered M1 and B1-3
  differ by less than one degree: $\alpha$, $\alpha'$=126$^\circ$, $\beta$,
  $\beta'$=245$^\circ$,  $\gamma$, $\gamma'$= 111$^\circ$, $\delta$=112$^\circ$,
  $\theta$=65$^\circ$. The different C-C bond length $d1$ in the two adducts
  of B2 lifts the symmetry between the two isotopically labeled nuclei and
  results into different HFI parameters [cf.\ B2(1) and B2(2) in Table \ref{table3}].
  }
  \end {center}
  \end {figure}

  \begin{table}[h!]
  \caption{\label{table3}
  Calculated HFI parameters of the fullerene derivatives M1 and B1-3 shown in 
  Fig.~\ref{fig1}. 
  We present data from pseudopotential calculations including core polarization 
  correction (PP-CP), and we also provide the calculated Fermi contact term
  without core polarization correction (PP) for completeness. 
  The dipolar terms are the same in both cases.
  For the fullerene mono-adduct M1 we also provide AE data and experimental data
  from Ref.~\onlinecite{marcus} (measured at 20 K) 
  for comparison. In the case of the B1-3 systems the values for the two 
  $^{13}$C-labeled nuclei are reported as (1) and (2), respectively. 
  All the data are in MHz.}
  \vspace{0.2cm}
  \begin{tabular}{clcr r r r r r r rr r r r}
  \hline
  \hline
  &Structure&   &$a$ &    &b$_{xx}$ &      &b$_{yy}$  &      &b$_{zz}$\\
  \hline
  &M1, PP-CP  &   & 2.0           &    & -1.6     &      & -2.3      &      & 4.0 \\
  &M1, PP    & & 4.1   &    &      &      &      &      &  \\
  &M1, AE     &   & 1.0           &    & -1.3     &      & -2.0      &      & 3.4 \\
  &M1, Expt.   &   & 2.0           &    &      &      &       &      &  \\
  \hline
  &B1(1), PP-CP      &   & 1.0           &    & -0.7     &      & -1.2      &      & 1.9 \\
&B1(1), PP      &   & 2.2           &    &      &      &       &      &  \\
  &B1(2), PP-CP     &   & 1.0           &    & -0.7     &      & -1.2      &      & 1.9 \\
  &B1(2), PP     &   & 2.2           &    &      &      &       &      &  \\
  \hline  
  &B2(1), PP-CP      &   & 0.2          &    & 0.0        &      & 0.6       &      & -0.6\\
&B2(1), PP      &   & 0.3           &    &      &      &       &      &  \\
  &B2(2), PP-CP     &   & 1.5          &    & -2.3     &      & -2.4      &      & 4.7 \\
  &B2(2), PP     &   & 4.4           &    &      &      &       &      &  \\
  \hline
  &B3(1), PP-CP     &   & 2.2           &    & -1.2    &      & -1.8      &      & 3.1 \\
&B3(1), PP      &   & 3.9           &    &      &      &       &      &  \\
  &B3(2), PP-CP     &   & 2.2           &    & -1.2    &      & -1.8      &      & 3.1 \\
  &B3(2), PP     &   & 3.9           &    &      &      &       &      &  \\
  \hline
  \hline
  \end{tabular}
  \label{tableIII}
  \end{table}

It is interesting to observe in Table \ref{table3} that the location of 
the diethyl malonate adduct has a significant effect on the HFI parameters,
leading to variations of up to an order of magnitude in the Fermi contact term.

For both the trans-1 isomer B1 and the trans-3 isomer B3 Table \ref{table3} shows 
that the two $^{13}$C nuclei have the same HFI tensor. At variance with this,
in the equatorial conformer B2 only one nucleus exhibits a sizeable HFI.
In order to rationalize this finding we examine the structural parameters
of each fullerene derivative. Figure.~\ref{fig4} shows the geometry of the
diethyl malonate adducts on the C$_{60}$ cage. The structure of the adducts
(i.e.\ bond lengths and bond angles) are almost identical in each fullerene
derivative. The only noticeable difference is that the length of the C-C bond
bridging the adduct to the fullerene is $1.58$~\AA\ for both adducts 
in the structures B1 and B3, while in the structure B2 the two adducts have different
C-C bond lengths of 1.60 and 1.57 \AA, respectively. This indicates that in
the equatorial conformer B2 one of the two nuclear spins is closer than the
other to the C$_{60}$ cage. Since the spin density is largest on the fullerene cage,
the nuclear spin with the shorter C-C bond length can be expected to exhibit a larger
HFI strength, consistent with the results in Table \ref{table3}.

In Ref.~\onlinecite{marcus} it was shown that, in systems like the structures
B1-3 considered here, it should be possible to achieve the entanglement 
of the two nuclear spins if the following conditions are fulfilled:
(i) the Fermi contact interaction when the
system is photo-excited to its triplet state is in the range 2-16~MHz.
(ii) the two isotopically
labeled C nuclei exhibit a very similar strength of the HFI.
Table \ref{table3} shows that, on the one hand, the trans-1 B1 and the equatorial 
B2 isomers do not fulfill at least one of these requirements. On the other hand,
the trans-3 B3 isomer fulfills both criteria, and hence it represents
a promising candidate for nuclear spin entanglement. 
This finding is particularly interesting since the B3 structure can be
produced with the relatively high yield of 12\% (cf.\ Sec.~\ref{sec.systems}).

\section{Conclusions}\label{sec.conc}

In this work we perform calculations of the hyperfine couling tensor for
photo-excited fullerene derivatives of interest in spin-based quantum 
information processing.
We report extensive validation of pseudopotential methods against all-electron
calculations and experimental data for a series of small radicals. A key
observation is that PP calculations can deliver an accuracy comparable
to AE calculations, provided the effect of core polarization is included.

We investigate the HFI strength in a series of photo-excited diethyl malonate fullerene
derivates, which are potential candidates for nuclear spin entanglement
operations. We clarify the relation between the calculated HFI parameters
and the underlying atomic-scale structure of these systems. We identify
one particular structure, the trans-3 isomer B3, as the best match to the
entanglement criteria set in Ref.~\onlinecite{marcus}. The fullerene
derivative thus identified can be produced with relatively high yield and thus represents
a potential candidate for achieving the entanglement of two nuclear spins 
in a photo-excited chromophore.

Our work provides a set of validation tests and benchmarks for future calculations
of HFI parameters using pseudopotential methods. In addition, this work represents
a first step towards the synergistic use of atomistic modelling 
and electron spin resonance experiments in the design 
of functional materials for quantum information technology.

\begin{acknowledgments}
We thank K. Porphyrakis, A. Ardavan and G. A. Briggs for stimulating discussions. 
The calculations were performed at the Oxford Supercomputing Centre. The spin
density images were rendered using Gabedit 2.3.0.\cite{gabedit}
We acknowledge support from the European Research Council under the European 
Community's Seventh Framework Programme (FP7/2007-2013) / ERC grant
agreements no. 239578 (F.G.) and 279781 (J.J.L.M.). JJLM thanks the
EPSRC (EP/I035536/1), the Royal Society, and St. John's College Oxford.
\end{acknowledgments}


\begin{thebibliography}{}%
\makeatletter
\providecommand \@ifxundefined [1]{%
 \ifx #1\undefined \expandafter \@firstoftwo
 \else \expandafter \@secondoftwo
\fi
}%
\providecommand \@ifnum [1]{%
 \ifnum #1\expandafter \@firstoftwo
 \else \expandafter \@secondoftwo
\fi
}%
\providecommand \enquote [1]{``#1''}%
\providecommand \bibnamefont  [1]{#1}%
\providecommand \bibfnamefont [1]{#1}%
\providecommand \citenamefont [1]{#1}%
\providecommand\href[0]{\@sanitize\@href}%
\providecommand\@href[1]{\endgroup\@@startlink{#1}\endgroup\@@href}%
\providecommand\@@href[1]{#1\@@endlink}%
\providecommand \@sanitize [0]{\begingroup\catcode`\&12\catcode`\#12\relax}%
\@ifxundefined \pdfoutput {\@firstoftwo}{%
 \@ifnum{\z@=\pdfoutput}{\@firstoftwo}{\@secondoftwo}%
}{%
 \providecommand\@@startlink[1]{\leavevmode\special{html:<a href="#1">}}%
 \providecommand\@@endlink[0]{\special{html:</a>}}%
}{%
 \providecommand\@@startlink[1]{%
  \leavevmode
  \pdfstartlink
   attr{/Border[0 0 1 ]/H/I/C[0 1 1]}%
   user{/Subtype/Link/A<</Type/Action/S/URI/URI(#1)>>}%
  \relax
 }%
 \providecommand\@@endlink[0]{\pdfendlink}%
}%
\providecommand \url  [0]{\begingroup\@sanitize \@url }%
\providecommand \@url [1]{\endgroup\@href {#1}{\urlprefix}}%
\providecommand \urlprefix [0]{URL }%
\providecommand \Eprint[0]{\href }%
\@ifxundefined \urlstyle {%
  \providecommand \doi [1]{doi:\discretionary{}{}{}#1}%
}{%
  \providecommand \doi [0]{doi:\discretionary{}{}{}\begingroup
  \urlstyle{rm}\Url }%
}%
\providecommand \doibase [0]{http://dx.doi.org/}%
\providecommand \Doi[1]{\href{\doibase#1}}%
\providecommand \bibAnnote [3]{%
  \BibitemShut{#1}%
  \begin{quotation}\noindent
    \textsc{Key:}\ #2\\\textsc{Annotation:}\ #3%
  \end{quotation}%
}%
\providecommand \bibAnnoteFile [2]{%
  \IfFileExists{#2}{\bibAnnote {#1} {#2} {\input{#2}}}{}%
}%
\providecommand \typeout [0]{\immediate \write \m@ne }%
\providecommand \selectlanguage [0]{\@gobble}%
\providecommand \bibinfo [0]{\@secondoftwo}%
\providecommand \bibfield [0]{\@secondoftwo}%
\providecommand \translation [1]{[#1]}%
\providecommand \BibitemOpen[0]{}%
\providecommand \bibitemStop [0]{}%
\providecommand \bibitemNoStop [0]{.\EOS\space}%
\providecommand \EOS [0]{\spacefactor3000\relax}%
\providecommand \BibitemShut [1]{\csname bibitem#1\endcsname}%
\end{thebibliography}%


\begin{thebibliography}{99}

\bibitem{morton}
J. J. L. Morton, A. M. Tyryshkin, R. M. Brown,
S. Shankar, B. W. Lovett, A. Ardavan, T. Schenkel,
E. E. Haller, J. W. Ager, and S. A. Lyon,
Nature {\bf 455}, 1085 (2008).

\bibitem{marcus}
M. Schaffry, V. Filidou, S. D. Karlen, E. M. Gauger, S. C. Benjamin,
H. L. Anderson, A. Ardavan, G. A. D. Briggs, K. Maeda, K. B. Henbest,
F. Giustino, J. J. L. Morton, and  B. W. Lovett,
Phys.\ Rev.\ Lett.\ {\bf 104}, 200501 (2010).

\bibitem{shift}
E. Zurek, C. J. Pickard, and J. Autschbach,
J.\ Phys.\ Chem.\ A\ {\bf 113}, 4117 (2009).

\bibitem{shift2}
M. d'Avezac, N. Marzari, and F. Mauri,
Phys.\ Rev.\ B\ {\bf 76}, 165122 (2007).

\bibitem{shift3}
L. Shao, J. R. Yates, and J. Titman,
J.\ Phys.\ Chem.\ A\ {\bf 111}, 13126 (2007).

\bibitem{shift4}
F. Mauri, B. G. Pfrommer, and S. G. Louie,
Phys.\ Rev.\ Lett.\ {\bf 77}, 5300 (1996).

\bibitem{shield}
L. Truflandier, M. Paris, and F. Boucher,
Phys.\ Rev.\ B\  {\bf76}, 035102 (2007).

\bibitem{shield2}
S. Rossano, F. Mauri, C. J. Pickard, and  I. Farnan,
J.\ Phys.\ Chem.\ B\ {\bf109}, 7245  (2005).

\bibitem{tens}
C. J. Pickard and F. Mauri,
Phys.\ Rev.\ Lett.\ {\bf88}, 086403 (2002).

\bibitem{tens2}
S. A. Joyce, J. R. Yates, C. J. Pickard, and S. P. Brown,
J.\ Am.\ Chem.\ Soc.\ {\bf 130}, 12663 (2008).

\bibitem{gten}
U. Gerstmann, M. Rohrmuller, F. Mauri, and W. G. Schmidt,
Physica\ Status\ Solidi\ {\bf 7}, 157 (2009).

\bibitem{gten2}
U. Gerstmann, A. P. Seitsonen, D. Ceresoli, F. Mauri, 
H. J. von Bardeleben, J. L. Cantin, and J. Garcia Lopez,
Phys.\ Rev.\ B\ {\bf 81}, 195208 (2010).

\bibitem{bio}
A.J. Hoff, in \textit{Advanced EPR: Application in Biology and 
Biochemistry}, (Elsevier, Amsterdam, 1989).

\bibitem{mat}
L. van Keand and  M. K. Bowman, in \textit{ Modern Pulsed and Continuous 
Wave Electron Spin Resonance}, (Wiley, New York, 1990).

\bibitem{blochl}
P. E. Bl\"ochl,
Phys.\ Rev.\ B\ {\bf 50}, 17953 (1994) 

\bibitem{vdw}
C. Van de Walle and P. E. Bl\"ochl,
Phys.\ Rev.\ B\ {\bf 47}, 4244 (1993).

\bibitem{hirsh}
A. Hirsch, I. Lamparth, and H. Karfunkel,
Angewandte Chemie.\ {\bf 33}, 437 (1994).

\bibitem{gga}
J. P. Perdew, K. Burke, and M. Ernzerhof,
Phys.\ Rev.\ Lett.\ {\bf 77}, 3865 (1996).

\bibitem{tm}
N. Troullier and J. L. Martins,
Phys.\ Rev.\ B {\bf 43}, 1993 (1991).

\bibitem{fuchs}
M. Fuchs and M. Scheffler,
Comput.\ Phys.\ Comm.\ {\bf 119}, 67 (1999).

\bibitem{gipaw}
C. Pickard and F. Mauri,
Phys.\ Rev.\ B\ {\bf 63}, 245101 (2001).

\bibitem{FW}
A. J. Freeman and R. E. Watson,
Phys.\ Rev.\ Lett.\ {\bf 5}, 489 (1960).

\bibitem{WF}
R. E. Watson and A. J. Freeman,
Phys.\ Rev.\ {\bf 123}, 2027 (1961).

\bibitem{corerelax}
M. S. Barhamy, M. H. F. Sluiter, and Y. Kawazoe,
Phys.\ Rev.\ B {\bf 76}, 035124 (2007).

\bibitem{qe}
P. Giannozzi, S. Baroni, N. Bonini, M. Calandra, R. Car, C. Cavazzoni, 
D. Ceresoli, G.L. Chiarotti, M. Cococcioni, I. Dabo, A. Dal Corso, 
S. Fabris, G. Fratesi, S. de Gironcoli, R. Gebauer, U. Gerstmann, 
C. Gougoussis, A. Kokalj, M. Lazzeri, L. Martin-Samos, N. Marzari, 
F. Mauri, R. Mazzarello, S. Paolini, A. Pasquarello, L. Paulatto, 
C. Sbraccia, S. Scandolo, G. Sclauzero, A. P. Seitsonen, 
A. Smogunov, P. Umari, and R. M. Wentzcovitch, 
J.\ Phys.\ Condens.\ Matter\ {\bf 21}, 395502 (2009); http://www.quantum-espresso.org 

\bibitem{gau}
Gaussian 09, Revision A.1, M. J. Frisch, G. W. Trucks, H. B. Schlegel, 
G. E. Scuseria, M. A. Robb, J. R. Cheeseman, G. Scalmani, V. Barone, 
B. Mennucci, G. A. Petersson, H. Nakatsuji, M. Caricato, X. Li, 
H. P. Hratchian, A. F. Izmaylov, J. Bloino, G. Zheng, J. L. Sonnenberg, 
M. Hada, M. Ehara, K. Toyota, R. Fukuda, J. Hasegawa, M. Ishida, 
T. Nakajima, Y. Honda, O. Kitao, H. Nakai, T. Vreven, J. A. Montgomery, 
Jr., J. E. Peralta, F. Ogliaro, M. Bearpark, J. J. Heyd, E. Brothers, 
K. N. Kudin, V. N. Staroverov, R. Kobayashi, J. Normand, K. Raghavachari, 
A. Rendell, J. C. Burant, S. S. Iyengar, J. Tomasi, M. Cossi, N. Rega, 
J. M. Millam, M. Klene, J. E. Knox, J. B. Cross, V. Bakken, C. Adamo, 
J. Jaramillo, R. Gomperts, R. E. Stratmann, O. Yazyev, A. J. Austin, 
R. Cammi, C. Pomelli, J. W. Ochterski, R. L. Martin, K. Morokuma, 
V. G. Zakrzewski, G. A. Voth, P. Salvador, J. J. Dannenberg, S. Dapprich, 
A. D. Daniels, Ö Farkas, J. B. Foresman, J. V. Ortiz, J. Cioslowski, 
and D. J. Fox, Gaussian, Inc., Wallingford CT, 2009.

\bibitem{eprII}
V. Barone, in \textit{Recent Advances in Density Functional Methods}, 
(World Scientific, Singapore, 1996), Part I.

\bibitem{trip1}
O. Gunnarsson and B. I. Lundqvist,
Phys.\ Rev.\ B\ {\bf 13}, 4274  (1976).

\bibitem{trip2}
Y. Kanai and J. C. Grossman,
Nano\ Letters {\bf 7}, 1967 (2007).

\bibitem{ch2ch13iso-a}
R. W. Fessenden and R. H. Schuler, 
J.\ Chem.\ Phys.\ {\bf 39}, 2147 (1963).

\bibitem{ch2ch13iso-b}
R. W. Fessenden,
J.\ Phys.\ Chem.\ {\bf 71}, 74 (1967).

\bibitem{co2iso}
D. W. Ovenall and D. H. Whiffen,
Mol.\ Phys.\ {\bf 4}, 135 (1961).

\bibitem{co3iso}
G. W. Chantry, A. Horsfield, J. R. Morton, and D. H. Whiffen,
Mol.\ Phys.\ {\bf 5}, 589 (1962).

\bibitem{h3coiso}
T. Momose, Y. Endo, E. Hirota, and T. Shida,
J.\ Chem.\ Phys.\ {\bf 88}, 5338 (1988).

\bibitem{h3co}
H. Nakatsuji, and M. Izawa,
J.\ Chem.\ Phys.\ {\bf 91}, 6205 (1989).

\bibitem{hcoiso-aniso}
R. W. Holmberg,
J.\ Chem.\ Phys.\ {\bf 51}, 3255 (1969).

\bibitem{h2coiso} 
L. B. Knight and J. Steadman,
J.\ Chem.\ Phys.\ {\bf 88}, 5338 (1988).

\bibitem{ch3iso} 
D. M. Chipman,
Theor.\ Chim.\ Acta.\ {\bf 82}, 93 (1992).

\bibitem{ch2chaniso}
M. Iwasaki, B. Eda, and K. Toriyama, 
J.\ Chem.\ Phys.\ {\bf 52}, 3837 (1970). 

\bibitem{ch2ch3aniso}
T. Shiga, H. Yamaoka, and A. Lund,
Z.\ Naturforsch.\ {\bf 29a}, 653 (1974).

\bibitem{h2coaniso}
L. B. Knight and J. Steadman,
J.\ Chem.\ Phys.\ {\bf 80}, 1018 (1984).

\bibitem{ch3anisoC}
R. W. Fessenden,
J.\ Phys.\ Chem.\ {\bf 71}, 74 (1967).

\bibitem{ch3anisoH}
J. Janecka, H. M. Vyas, and M. Fujimoto,
J.\ Chem.\ Phys.\ {\bf 54}, 3229 (1971).

\bibitem{ext1}
D. J. van den Heuvel {\it et al.}, 
Chem.\ Phys.\ Lett.\ {\bf 23}, 18l (1994).

\bibitem{ext2}
A. Sassara, G. Zerza, and M. Chergui,
Chem.\ Phys.\ Lett.\ {\bf 261}, 213 (1996).

\bibitem{gabedit}
A. Alouche, http://gabedit.sourceforce.net/.

\end{thebibliography}
\end{document}